\documentclass[aps,prb,twocolumn,superscriptaddress,showpacs]{revtex4}

\usepackage{amsmath}    % need for subequations
\usepackage{graphicx}   % need for figures
\usepackage{verbatim}   % useful for program listings
\usepackage{color}      % use if color is used in text
\usepackage{subfigure}  % use for side-by-side figures
\usepackage{hyperref}   % use for hypertext links, including those to external documents and URLs

\usepackage{times}

\begin{document}

\title{Nature of the low temperature ordering of Pr in PrBa$_{2}$Cu$_{3}$O$_{6+x}$}

\author{Annam\'{a}ria Kiss}
\email{akiss@szfki.hu}
\affiliation{Budapest University of Technology and Economics, Institute
of Physics and Condensed Matter Research Group of the Hungarian Academy of
Sciences, H-1521 Budapest, Hungary}

\author{Ferenc Simon}
\affiliation{Budapest University of Technology and Economics, Institute
of Physics and Condensed Matter Research Group of the Hungarian Academy of
Sciences, H-1521 Budapest, Hungary}

\affiliation{Universit\"{a}t Wien, Fakult\"{a}t f\"{u}r Physik, Strudlhofgasse 4, 1090 Wien, Austria}

\begin{abstract}

Theoretical model is presented to describe the anomalous ordered phase of Pr ions in PrBa$_{2}$Cu$_{3}$O$_{6+x}$ below $T_{\rm Pr} \approx 12-17$\,K.  The model considers the Pr multipole degrees of freedom and coupling between the Cu and Pr subsystems.
We identify the symmetry allowed coupling of Cu and Pr ions and conclude that only an $ab$-plane Pr dipole ordering can explain the Cu spin rotation observed at $T_{\rm Pr}$ by neutron diffraction by Boothroyd \textit{et al.} [A. T. Boothroyd \textit{et al.}, Phys. Rev. Lett. {\bf 78}, 130 (1997)].
A substantial enhancement of the Pr ordering temperature is shown to arise from the Cu-Pr coupling which is the key for the anomalous magnetic behavior in PrBa$_{2}$Cu$_{3}$O$_{6+x}$.

\end{abstract}

\maketitle

\section{Introduction}

Non-superconducting PrBa$_{2}$Cu$_{3}$O$_{6+x}$ compound attracted considerable attention in the last two decades because of its intriguing magnetic and electronic properties within the ReBa$_2$Cu$_3$O$_{6+x}$ family of compounds (Re=rare earth atom).\cite{radousky-1992} In spite of the great efforts from both experimental and theoretical sides, numerous problems remain unresolved such as the suppression of superconductivity and the nature of the long-range ordered state of Pr sublattice with a unique, about an order of magnitude larger than for other rare earth atoms, ordering temperature.
Although, there exists a discussion whether the ground state of PrBa$_{2}$Cu$_{3}$O$_{6+x}$ is really a non-superconducting, insulating material with a magnetically ordered Cu and Pr sublattices, herein we consider this modification of PrBa$_{2}$Cu$_{3}$O$_{6+x}$ as "canonical" and treat it solely herein.
We note that superconductivity in PrBa$_{2}$Cu$_{3}$O$_{6+x}$ was reported\cite{Zou-1998} and a non-magnetic Pr ground state was found by $^{141}$Pr NMR.\cite{nehrke-1996}
It was suggested that these observations may result from a non-stoichiometric compound, i.e. when Pr occupies only about a half of the rare-earth sites, and the other half is occupied by the non-magnetic Ba.\cite{narozhnyi-1999}

The theoretical model of Fehrenbacher and Rice, i.e. the hybridization of Pr $4f$ and the nearest neighbor O $2p$ orbitals is the most accepted model to account for the absence of superconductivity in PrBa$_{2}$Cu$_{3}$O$_{6+x}$. It is suggested that the localization of holes in the hybridized $4f$-$2p$ orbitals renders the material non-superconducting.\cite{fehrenbacher-1993}
Although the non-superconducting nature of PrBa$_{2}$Cu$_{3}$O$_{7}$ compound is a challenging and interesting problem, we concentrate herein exclusively on the low-temperature Pr ordering.

PrBa$_{2}$Cu$_{3}$O$_{6+x}$ is an insulator for every $x$ and the Cu spins in the CuO$_{2}$ planes order antiferromagnetically at temperatures of $T_{\rm N}\sim 350$\,K and $250$\,K for $x=0$ and $x=1$, respectively.\cite{cooke-1990, felner-1989, boothroyd}
As the temperature is further decreased, the Pr sublattice also undergoes a phase transition\cite{kebede-1989, li-1989} in the temperature range $T_{\Pr}\sim 12 - 17$\,K depending on $x$, which appears as an anomaly in the temperature dependence of thermodynamic quantities.\cite{hilscher-1994, longmore}
M\"ossbauer spectroscopy,\cite{hodges-1993} neutron diffraction,\cite{boothroyd, hilscher-1994, longmore} and NMR\cite{nehrke-1996, staub-1996} showed that the Pr magnetic moments order antiferromagnetically below $T_{\Pr}$. However, there is no consensus among these experiments with respect to the magnitude of the ordered Pr moment, its direction, and the nature of this transition.

Neutron diffraction studies found $0.56\mu_{B}$ ($x=0.92$) and $1.15\mu_{B}$ ($x=0.35$) ordered moment with direction tilted out from the $ab$-plane.\cite{boothroyd} The rotation of Cu moments within the CuO$_{2}$ plane was found to accompany the Pr ordering,\cite{boothroyd, lister-2001} which was suggested to result from the strong coupling between the Pr and Cu subsystems. However, the magnitude of the coupling, its character, and its influence on the magnitude of the ordered Pr magnetic moment, and for the ordering temperature is yet unexplained.

We give herein a comprehensive description of the Pr ordering in PrBa$_{2}$Cu$_{3}$O$_{6}$ on the basis of a model of localized $4f^{2}$ electrons of the Pr ion.
We construct the general form of the Cu-Pr interaction allowed by the symmetry and identify the Pr order parameter from the rotation of Cu spins below $T_{\rm Pr}$ observed by neutron diffraction.\cite{boothroyd, lister-2001}
Although pseudo-dipole Cu-Pr interaction has been already proposed,\cite{boothroyd, lister-2001, maleev} we derive the Cu-Pr interaction in general including all Pr multipole moments and arbitrary wave vectors.
We describe a crystalline electric field (CEF) model which takes a ground state quasi-triplet composed by a doublet and a singlet states and also includes coupling between the Pr and Cu subsystems. This low-energy scheme is consistent with the neutron diffraction results\cite{hilscher-1994} and also with the high-temperature susceptibility data.\cite{uma} We also discuss the enhancement of the Pr ordering temperature due to the Cu-Pr coupling and the magnitude of Pr magnetic moment in the ordered state.

\section{General form of the Cu-Pr interaction}\label{cu-pr-interaction}

\begin{figure}
\centering
\includegraphics[totalheight=6.2cm,angle=0]{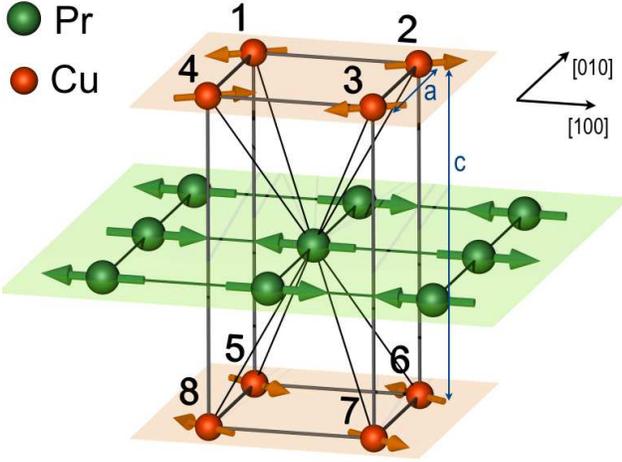}
\caption{Magnetic structure of Cu spins in the CuO$_{2}$ planes and Pr dipole moments in the $ab$-plane in the temperature range $T \le T_{\rm Pr} $.
 The central Pr ion in the $ab$-plane is located at position ${\bf r}_{0}=(0,0,0)$, and the numbers label the Cu sites, i.e. Cu[1-8] used in the text.}\label{cu-pr-fig}
\end{figure}

In PrBa$_{2}$Cu$_{3}$O$_{6}$,
the Cu spins are antiferromagnetically ordered along the $[100]$ direction in the CuO$_{2}$ planes with the ordering vector ${\bf Q}=[\pi/a, \pi/a, \pi/c]$ below the
temperature $T_{\rm Cu}\approx 350$\,K (phase AFI).\cite{boothroyd}
As the temperature is further decreased, the Pr sublattice also undergoes an ordering at temperature $T_{\rm Pr}= 12$\,K.
This ordering is accompanied by the rotation of Cu spins within the CuO$_{2}$ planes in such a way that the new spin structure is non-collinear along the $c$-direction (phase AFIII).\cite{boothroyd, lister-2001, hill} The Cu spin rotation is characterized by the ordering vector $\widetilde{\bf Q}=[\pi/a, \pi/a, 0]$.
The Cu spin structure in the CuO$_{2}$ planes is shown in Fig.~\ref{cu-pr-fig} in the temperature range $T \le T_{\rm Pr}$.

First, we construct the invariant form of the interaction between the multipole moments of a Pr ion at position ${\bf r}_{0}=(0,0,0)$ and the surrounding eight Cu spins, Cu[1-8], shown in Fig.~\ref{cu-pr-fig}.
Although the local symmetry at Pr site is tetragonal, a Cu-Pr pair has lower symmetry.
For example, the Cu[1]-Pr pair has reflection symmetry with respect to the $[1,1,0]$ mirror plane.
The transformation of total angular momentum ${\bf J}$ of Pr ion under this operation is: $J_{x} \rightarrow J_{y}$, $J_{y} \rightarrow J_{x}$, and $J_{z} \rightarrow -J_{z}$. The same transformation is applied for the Cu spin ${\bf S}$. Thus, spin operators $J_{x}-J_{y}$ ($S_{x}-S_{y}$) and $J_{z}$ ($S_{z}$) are odd, while $J_{x}+J_{y}$ ($S_{x}+S_{y}$) is even under this reflection.
Not only the bilinear products of the Pr dipole moments and Cu spins appear in the interaction but also the bilinear products of rank-3 octupole and rank-5 triakontadipole\cite{kuramoto-2009} operators of Pr ion and Cu spins since they are allowed by the time-reversal symmetry.
The octupole operators $T^{\beta}_{z}$ and $T^{\beta}_{x}-T^{\beta}_{y}$ are even, while $T_{xyz}$ and $T^{\beta}_{x}+T^{\beta}_{y}$ are odd operators under the present transformation.
In the case of tetragonal symmetry, there is one more independent magnetic operator, namely the rank-5 triakontadipole operator $V_{1u}=\overline{J_{x}J_{y}J_{z}(J_{x}^2-J_{y}^2)}$ which is even under the present reflexion.
Thus,
the invariant form of the interaction for the Cu[1]-Pr pair is constructed as
\begin{eqnarray}
{\cal H}^{\rm Cu-Pr}_{I} [1] &=& (S_{x}+S_{y})\left[ c_{11}(J_{x}+J_{y}) + c_{12} T^{\beta}_{z} + c_{13}A_{1u}    \right] \nonumber\\
&+&   (S_{x}-S_{y})\left[ c_{21}(J_{x}-J_{y}) + c_{22} J_{z} + c_{23} T_{xyz}     \right] \nonumber\\
&+&
S_{z}\left[ c_{31}(J_{x}-J_{y}) + c_{32} J_{z} + c_{33} T_{xyz}     \right],
 \label{int1}
\end{eqnarray}
where $c_{ij}$ are constants which are not determined by the symmetry itself.
The pair Cu[2]-Pr is connected to the pair Cu[1]-Pr by a $\pi/2$ rotation around the $c$-axis. Under this operation the transformation of the angular momentum components is given by $J_{x} \rightarrow -J_{y}$, $J_{y} \rightarrow J_{x}$, and $J_{z} \rightarrow J_{z}$. The same transformation holds also for the Cu spin components.
Thus, the form of the interaction for the Cu[2]-Pr pair can be obtained from Eq.~(\ref{int1}) by applying the $\pi/2$ rotation.
Repeating further the appropriate rotations and reflections, the interaction can be obtained for all the eight Cu-Pr pairs.
The Cu spin at position ${\bf r}$ is expressed by Fourier transformation as
${\bf S}({\bf r}) = \sum_{\bf q}  {\bf S}^{\bf q} {\rm e}^{i {\bf q} {\bf r}}$. Using this expression, we obtain the total interaction as
\begin{widetext}
\begin{eqnarray}
{\cal H}^{\rm Cu-Pr}_{I} &=& \sum_{k=1}^{8}  {\cal H}^{\rm Cu-Pr}_{I} [k] =  8\sum_{\bf q} \left\{
(c_{11} + c_{21})(S_{x}^{\bf q}J_{x}+S_{y}^{\bf q}J_{y})c_{x}c_{y}c_{z}  +  (c_{11} - c_{21})(S_{x}^{\bf q}J_{y}+S_{y}^{\bf q}J_{x})s_{x}s_{y}c_{z} \right. \nonumber\\
&+& \left.
c_{22}J_{z} \left[ S_{x}^{\bf q} s_{x} c_{y} s_{z} - S_{y}^{\bf q} c_{x} s_{y} s_{z}  \right]
 +  c_{31} \left[ J_{x} S_{z}^{\bf q} s_{x} s_{y} c_{z} - J_{y} S_{z}^{\bf q} c_{x} s_{y} s_{z} \right] 
 + c_{32} J_{z} S_{z}^{\bf q} c_{x} c_{y} c_{z}
\right. \nonumber\\
&+& \left.
 c_{12} T^{\beta}_{z} \left[ S_{x}^{\bf q} s_{x} c_{y} s_{z} + S_{y}^{\bf q} c_{x} s_{y} s_{z}  \right]
 +
 c_{13} A_{1u} \left[ S_{x}^{\bf q} c_{x} s_{y} s_{z} + S_{y}^{\bf q} s_{x} c_{y} s_{z}  \right] +  c_{23} T_{xyz}  \left[ S_{x}^{\bf q} c_{x} s_{y} s_{z} - S_{y}^{\bf q} s_{x} c_{y} s_{z}  \right]
\right. \nonumber\\
&+& \left.
  c_{33} T_{xyz} S_{z}^{\bf q} s_{x} s_{y} c_{z}
   \right\}, \label{int}
\end{eqnarray}
\end{widetext}
where $c_{k}(s_{k})$ denotes  $\cos (q_{k}a/2) (\sin (q_{k}a/2))$ for  $k$=$x$, $y$, and $\cos (q_{k}c/2) (\sin (q_{k}c/2))$ for $k$=$z$.

Now we discuss the multipole order of Pr sublattice based on expression (\ref{int}).
In phase AFI, the Cu spin component $S_{x}^{\bf q}$ with ${\bf q}={\bf Q}$ is non-zero.
In this phase there is no coupling between the Cu spins and Pr magnetic moments because the wave vector ${\bf Q}$ leads to $c_{x}=c_{y}=c_{z}=0$ and $s_{x}=s_{y}=s_{z}=1$, which gives that none of the terms survives in expression (\ref{int}).
On the other hand, a coupling of Cu spins and Pr moments is present in phase AFIII.
The non-collinear Cu spin structure in phase AFIII can be described by the appearance of the extra spin component $S_{y}^{\widetilde{\bf Q}}$
in addition to the component $S_{x}^{{\bf Q}}$.
The wave vector $\widetilde{\bf Q}$ gives $c_{x}=c_{y}=0$, $c_{z}$=1, and $s_{x}=s_{y}=1$, $s_{z}=0$ in expression (\ref{int}). Together with the condition $ S_{y}^{\widetilde{\bf Q}} \ne 0$, we find that the only surviving term in the form of Cu-Pr interaction is the coupling $S_{y}^{\widetilde{\bf Q}} J_{x}$.
We note that the coupling $S_{x}^{\widetilde{\bf Q}} J_{y}$ is also allowed by symmetry since the $[100]$ and $[010]$ domains in phase AFI are equivalent, which gives that
the domains $\langle S_{y}^{\widetilde{\bf Q}} \rangle \ne 0, \langle S_{x}^{\widetilde{\bf Q}} \rangle = 0$ and $\langle S_{x}^{\widetilde{\bf Q}} \rangle \ne 0, \langle S_{y}^{\widetilde{\bf Q}} \rangle = 0$ are also equivalent.

Thus, we conclude that the neutron diffraction result for the Cu spin rotation below $T_{\rm Pr}$ is compatible with Pr dipole moments lying along the crystalline $[100]$ or $[010]$ direction, and no deviation from this direction is allowed by symmetry in contrast to the structure proposed in Ref.~\onlinecite{boothroyd}.
Since the $ S_{y}^{\widetilde{\bf Q}}(S_{x}^{\widetilde{\bf Q}})$ Cu spin components alternate along the $a$ and $b$ directions, the
Pr dipole moments also form an AFM structure in the $ab$-planes with the ordering vector $\widetilde{\bf Q}$.
The magnetic structure for the Cu spins and Pr dipole moments below $T_{\rm Pr}$ has the structure shown in
Fig.~\ref{cu-pr-fig}.\endnote{In our convention
$S_{y(x)}^{\widetilde{\bf Q}} =  -|S_{y(x)}^{\rm Cu}|$, where $|S_{y(x)}^{\rm Cu}|$ is the magnitude of the Cu spin at a position in the CuO$_{2}$ planes. Furthermore, $J_{x(y)}^{{\rm Pr},\widetilde{\bf Q}}=-|J_{x(y)}^{\rm Pr}|$, where $|J_{x(y)}^{\rm Pr}|$ is the magnitude of the Pr dipole moment in the $ab$-plane.}

\section{CEF model for the coupled Cu-Pr system}

We now turn to discuss a CEF model of the Pr ion $4f$ electrons to describe the Pr ordering in PrBa$_{2}$Cu$_{3}$O$_{6}$. The model consists of the coupled Pr and Cu subsystems and allows to calculate experimentally relevant quantities.
The Pr ion has 4$f^{2}$ electronic configuration in PrBa$_{2}$Cu$_{3}$O$_{6}$, which gives $J=4$ as the total angular momentum based on the Hund's rule for the ground state.
In the Pr subsystem, we consider AFM interaction between the in-plane  $J_{x}$ and $J_{y}$ dipole moments described by the Hamiltonian
\begin{eqnarray}
{\cal H}_{\rm Pr} &=& \frac{1}{2z}
\sum_{\langle i,j \rangle} \Lambda \left[ J_{x,i}J_{x,j} + J_{y,i}J_{y,j} \right],
\label{ham-pr}
\end{eqnarray}
where $z=4$ is the number of nearest-neighbor Pr ions and $\Lambda$ is the coupling constant. The Hamiltonian (\ref{ham-pr}) is taken in a quasi-triplet subspace, where the ground state is the doublet
\begin{eqnarray}
| d_{\pm} \rangle = a | \pm 3 \rangle + \sqrt{1-a^2} | \mp 1 \rangle \label{cef-doublet}
\end{eqnarray}
 with a near singlet excited state
 \begin{eqnarray}
| s \rangle = \frac{1}{\sqrt{2} } \left(  | 2 \rangle + | -2 \rangle \right),\label{cef-singlet}
\end{eqnarray}
where the states are expressed by the eigenstates of ${\bf J}$.
The doublet and singlet states are separated by the gap $\Delta$.\cite{hilscher-1994, uma}
The other CEF levels lie at much higher energy therefore these can be neglected when describing the low temperature behavior.

We treat the Cu subsystem phenomenologically and consider the free energy expansion:
\begin{eqnarray}
{\cal F}_{\rm Cu} &=& -\frac{1}{\beta}{\rm ln}{\rm Tr}_{S} \left({\rm e}^{-\beta {\cal H}_{\rm Cu}} \right)
\nonumber\\
&=& a_{s} ( \langle S_{x}^{\widetilde{\bf Q}} \rangle^2 + \langle S_{y}^{\widetilde{\bf Q}} \rangle^2  ) + b_{s} ( \langle S_{x}^{\widetilde{\bf Q}} \rangle^2 + \langle S_{y}^{\widetilde{\bf Q}} \rangle^2  )^2 \nonumber\\
&+& 4c_{s} \langle S_{x}^{\widetilde{\bf Q}} \rangle^2 \langle S_{y}^{\widetilde{\bf Q}} \rangle^2
+ {\cal F}_{\rm Cu}^{0},\label{free-cu}
\end{eqnarray}
where ${\cal F}_{\rm Cu}^{0}$ contains the part corresponding to the Cu spin components, $S_{x}^{\bf Q}$ and $S_{y}^{\bf Q}$, which are not relevant for our discussion. We assume that $\langle S_{y(x)}^{\widetilde{\bf Q}} \rangle$ is not critical around $T=T_{\rm Pr}$, thus, $a_{s}>0$. Parameters $b_{s}$ and $c_{s}$ determine the anisotropy as we discuss below.

The local coupling between the Cu and Pr subsystems is assumed as
\begin{eqnarray}
{\cal H}_{\rm c} =  z^{\prime} \sum_{i}\lambda \left[  J_{x,i} \langle S_{y}^{\widetilde{\bf Q}} \rangle
+ J_{y,i} \langle S_{x}^{\widetilde{\bf Q}} \rangle    \right],
\end{eqnarray}
which corresponds to the only non-vanishing term in expression~(\ref{int}) for $\widetilde{\bf Q}=[\pi/a, \pi/a, 0]$.
Here, $z^{\prime} = 8$ is the number of Pr nearest-neighbor Cu ions and $\lambda$ is the Cu-Pr coupling constant.
The series expansion of the total free energy has the form
\begin{eqnarray}
{\cal F} &=&  -\frac{1}{\beta}{\rm ln}{\rm Tr}_{S} \left({\rm e}^{-\beta {\cal H}_{\rm Cu}} \right)-\frac{1}{\beta}{\rm ln}{\rm Tr}_{J} \left({\rm e}^{-\beta \left({\cal H}_{\rm Pr}+{\cal H}_{\rm c}\right)} \right) \nonumber\\
& \equiv & {\cal F}_{0} + {\cal F}_{\rm Cu}^{0},
\label{free-whole}
\end{eqnarray}
where
\begin{eqnarray}
{\cal F}_{0}  &=&
a_{J} ( \langle J^{\rm Pr}_{x} \rangle^2  + \langle J^{\rm Pr}_{y} \rangle^2  ) + b_{J} ( \langle J_{x}^{\rm Pr} \rangle^2 
\nonumber\\
&+& \langle J_{y}^{\rm Pr} \rangle^2)^2 +
  \lambda z^{\prime}  (\langle  J_{x}^{\rm Pr} \rangle \langle S_{y}^{\widetilde{\bf Q}} \rangle
  + \langle J_{y}^{\rm Pr} \rangle \langle S_{x}^{\widetilde{\bf Q}} \rangle ) \nonumber\\
  &+& a_{s} ( \langle S_{x}^{\widetilde{\bf Q}} \rangle^2 + \langle S_{y}^{\widetilde{\bf Q}} \rangle^2  ) + b_{s} ( \langle S_{x}^{\widetilde{\bf Q}} \rangle^2 + \langle S_{y}^{\widetilde{\bf Q}} \rangle^2  )^2 \nonumber\\
&+& 4c_{s} \langle S_{x}^{\widetilde{\bf Q}} \rangle^2 \langle S_{y}^{\widetilde{\bf Q}} \rangle^2.
\label{free-0}
\end{eqnarray}
The coefficient $a_{J}$ is expressed in Eq.~(\ref{free-0}) as
\begin{eqnarray}
a_{J} = \frac{1}{2} \Lambda^2 \left( \frac{1}{\Lambda} + \frac{1}{\Delta}\frac{(2\gamma)^2({\rm e}^{-\Delta/T}-1)}{({\rm e}^{-\Delta/T}+2)} \right),
\end{eqnarray}
where $\gamma=1/4(9-2a^2+6\sqrt{7}a\sqrt{1-a^2})$ is the matrix element of the dipole operator, $J_{x}$, between the doublet and singlet states.

The quasi-triplet subspace does not carry anisotropy with respect to the dipole moment components $J_{x}$ and $J_{y}$, thus the Pr magnetic moment is isotropic in the $ab$-plane for the Pr subsystem. However, there is a weak anisotropy even for the structurally tetragonal $x=0$ compound which gives rise to a degenerate [100] and [010] easy axis ordering of the Cu spins in the AFI phase.\cite{JanossyPRB1998} We introduce this anisotropy of the Cu sublattice through the parameters $b_{s}$ and $c_{s}$ in expression~(\ref{free-cu}), which produces anisotropy also for the in-plane Pr magnetic moments through the Cu-Pr coupling. In the AFIII phase, these easy axes are inherited and the Cu spin ordering with wave vector $\widetilde{{\bf Q}}$ is reproduced when $c_s>0$. We assume an infinitesimally small magnetic field along the $[010]$ direction on the Cu sublattice, which resolves the [100], [010] degeneracy and stabilizes the solution $\langle J_{x} \rangle \ne 0$, $\langle J_{y} \rangle = 0$ against the solution $\langle J_{x} \rangle = 0$, $\langle J_{y} \rangle \ne 0$. The resulting Pr dipole moments have also a $[100]$ easy axis.

In our scenario, the Pr dipole moments order at $T=T_{\rm Pr}$, and the order $\langle J_{x} \rangle \ne 0$ induces the $\langle S_{y}^{\widetilde{\bf Q}} \rangle$ Cu spin component. From the condition $\partial {\cal F}/\partial \langle S_{y}^{\widetilde{\bf Q}} \rangle = 0$ we obtain\begin{eqnarray}
\langle S_{y}^{\widetilde{\bf Q}} \rangle = -\frac{\lambda z^{\prime}}{2a_{s}} \langle J^{\rm Pr}_{x} \rangle.\label{es-T}
\end{eqnarray}
Substituting this expression into the free energy expansion ${\cal F}$, the second order coefficient of $ \langle J^{\rm Pr}_{x} \rangle $ is obtained as
\begin{eqnarray}
\widetilde{a}_{J} \equiv a_{J} - \frac{(\lambda z^{\prime})^2}{4a_{s}}.
\end{eqnarray}

We define the temperature $T_{0}$ which corresponds to the non-coupled system with $\lambda=0$ and the temperature $T_{\rm Pr}$ for the coupled system $\lambda\ne 0$.
$T_{0}$ is obtained from the condition $a_{J} =0$, where the second-order coefficient has the form $a_{J} \equiv a_{J}^{\prime} (T-T_{0})$ in the vicinity of the phase transition. However, this transition temperature is modified due to the coupling to the Cu subsystem ($\lambda \ne 0$). $T_{\rm Pr}$ is obtained from the condition $\widetilde{a}_{J}=0$, which gives
\begin{eqnarray}
T_{\rm Pr} = T_{0} + \frac{(\lambda z^{\prime})^2}{4a_{J}^{\prime} a_{s}}.\label{trans-t}
\end{eqnarray}
Thus the Pr transition temperature is enhanced due to the Cu-Pr coupling irrespective of the sign of the coupling parameter $\lambda$.

The rotation of the Cu spins in the CuO$_{2}$ plane is expressed as
$\phi = {\rm tan}^{-1}(\langle S_{y}^{\widetilde{\bf Q}} \rangle/\langle S_{x}^{{\bf Q}} \rangle)$, where $\phi=0$ corresponds to the $x$-direction. The Cu magnetic moment is expressed as $\mu_{\rm Cu} = g_{\rm Cu} \mu_{\rm B} [\langle S_{y}^{\widetilde{\bf Q}} \rangle^2 + \langle S_{x}^{{\bf Q}} \rangle^2]^{1/2} $, where $g_{\rm Cu} \approx 2$ is the Cu g-factor. An ordered magnetic moment of $0.66\mu_{\rm B}$/Cu[2] is found also for the AFI phase due to quantum fluctuations.\cite{Manousakis1991} Our model includes this effect phenomenologically as this ordered magnetic moment is retained above $T_{\rm Pr}$ in the AFI phase. 

\begin{figure}
\centering
\includegraphics[totalheight=6.2cm,angle=0]{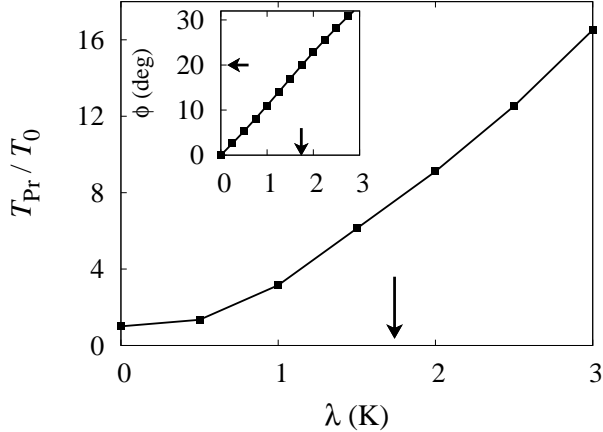}
\caption{Enhancement of the transition temperature, $T_{\rm Pr}/T_{0}$, due to the coupling parameter $\lambda$.
{\sl Inset} shows the Cu spin-rotation angle $\phi$ at $T=0$ as a function of $\lambda$.
The used parameter values are $\Delta=4$\,K, $\Lambda=1.2$\,K, $a=0.96$, and $a_{s}=140$. Arrows show the experimental values for $\phi$ and $\lambda$ from Refs.~\onlinecite{boothroyd} and ~\onlinecite{lister-2001}, respectively.}
\label{fig-tc-l}
\end{figure}

Figure~\ref{fig-tc-l} shows the enhancement of the Pr transition temperature due to the Cu-Pr coupling and the corresponding Cu spin rotation angle as a function of the coupling parameter $\lambda$, calculated within the above model.
We used $\Delta=4$\,K and $a=0.96$ for the CEF parameters, where the latter value is obtained from the high-temperature susceptibility data\cite{uma} in PrBa$_{2}$Cu$_{3}$O$_{6}$. Neutron scattering found the interaction parameters as $\Lambda=1.2$\,K and $\lambda=1.74$\,K\cite{lister-2001} and an ordered Cu moment of $\mu_{\rm Cu}=0.64\mu_{\rm B}$ for $x=0.35 $,\cite{boothroyd} which were used in the calculation of Fig.~\ref{fig-tc-l} as fix parameters. Clearly, the model accurately reproduces the $\phi = 20^{o}$ rotation of the Cu spins which was observed experimentally in Ref.~\onlinecite{boothroyd}. In addition, the model accounts for about 70 \% enhancement of the Pr ordering temperature with this parameter set.

\begin{figure}
\centering
\includegraphics[totalheight=6.4cm,angle=0]{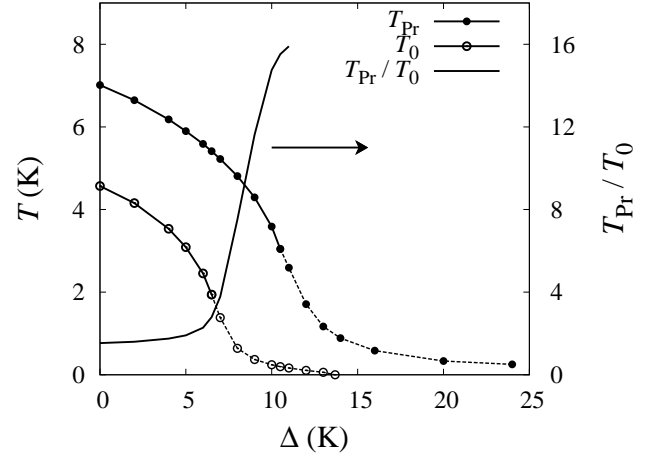}
\caption{Transition temperatures $T_{0}$ and $T_{\rm Pr}$ as a function of $\Delta$. The parameter values are chosen as $\Lambda=1.2$\,K, $\lambda=1.74$\,K, $a=0.96$. {\sl Solid line} corresponds to second-order, while {\sl dashed line} to first-order phase transition. The ratio $T_{\rm Pr}/T_{0}$ is also shown as continuous line.}\label{fig-tc-d}
\end{figure}

The ground state Pr magnetic moment and the enhancement of $T_{\rm Pr}$ strongly depends on the value of the CEF energy gap, $\Delta$.
In Fig.~\ref{fig-tc-d}, we show the calculated $T_{\rm Pr}$ as a function of $\Delta$, where we
fix the interaction and CEF parameters as $\Lambda=1.2$\,K, $\lambda=1.74$\,K, $a=0.96$. The value of the phenomenological parameter $a_{s}$ is chosen for each $\Delta$ value so that it reproduces the experimental values $\phi = 20^{o}$ and  $\mu_{\rm Cu}=0.64\mu_{\rm B}$.
For small values of $\Delta$ the phase transition is second-order. The transition temperatures $T_{0}$ and $T_{\rm Pr}$ are suppressed with increasing energy gap $\Delta$, and the phase transition changes to first-order above a critical value of $\Delta$. Increasing further $\Delta$, the phase transition disappears.
These behaviors are due to the fact that the $J_{x}$ dipole order is interaction-induced within the quasi-triplet subspace since the dipole operator $J_{x}$ has non-vanishing matrix element only between the doublet and singlet states. Around $\Delta \sim 10$\,K the ratio $T_{\rm Pr}/T_{0}$ is considerably enhanced due to the Cu-Pr coupling.

\section{Measurement of the local fields}

Finally, we discuss the field-angle dependence of the local magnetic field acting on a Pr ion.
This local field could be measured by a local magnetic probe such as e.g. $^{89}$Y using NMR\cite{alloul-1989} or Gd$^{3+}$ using
ESR\cite{janossy-1994} which can be substituted into the Pr sublattice in a low concentration.

\begin{figure}
\centering
\includegraphics[totalheight=5.5cm,angle=0]{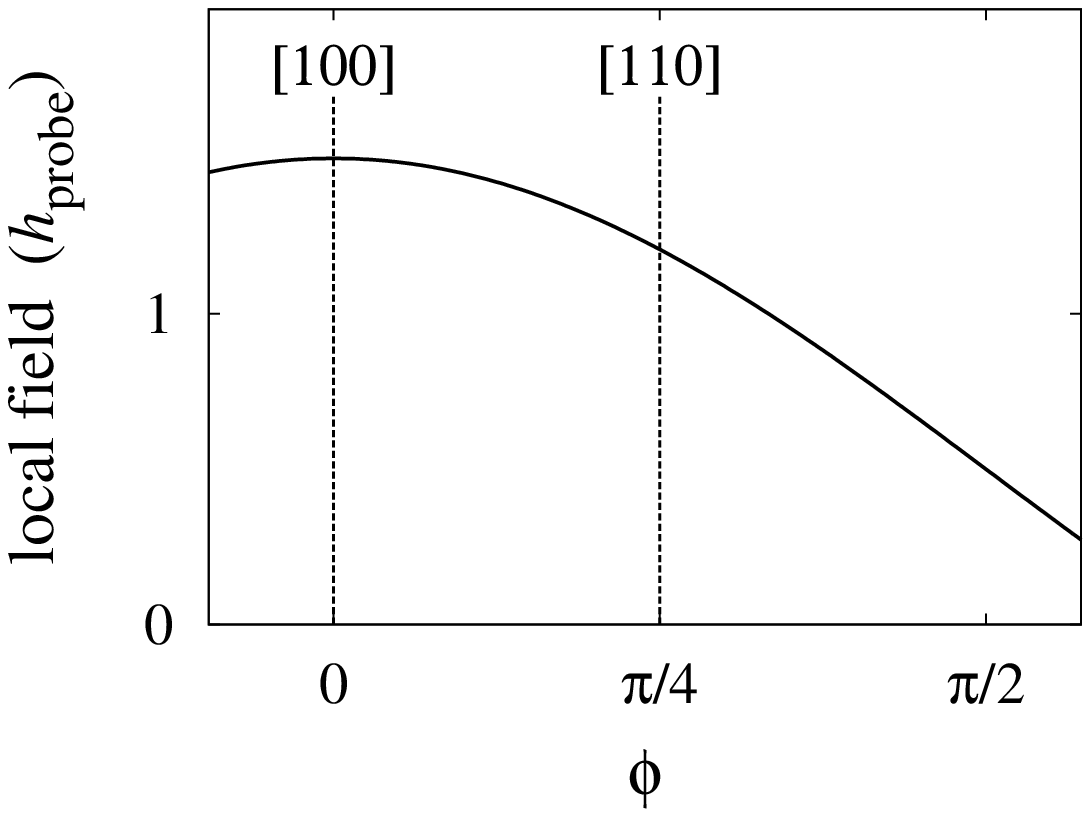}\\
\includegraphics[totalheight=5.5cm,angle=0]{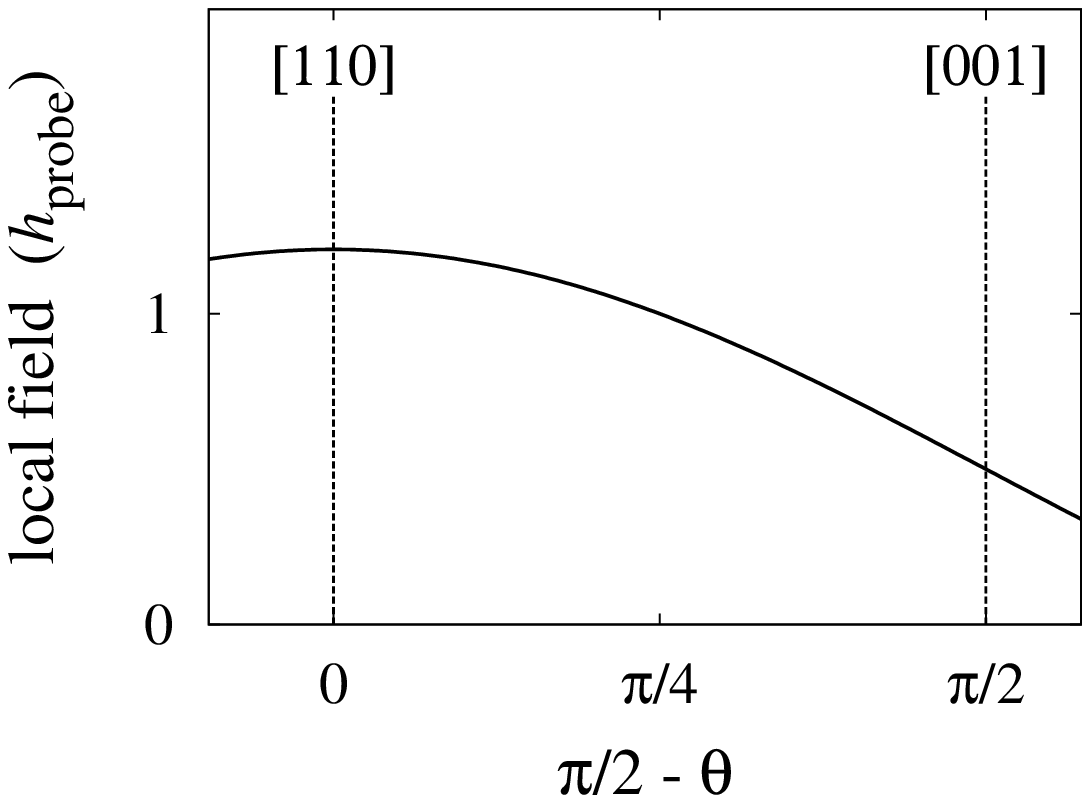}
\caption{Field-angle dependence of the local field $h_{\rm probe}$ acting on the local magnetic probe by changing the direction of the external magnetic field in the $[001]$ ({\sl top}) and $[110]$ ({\sl bottom}) planes with $\theta = \pi/2$ and $\phi=\pi/4$, respectively. In the plot we take the exemplifying values $h_{x}=1$ and $H=0.5$. The field angles are defined in the text.
}\label{pheno-angle-fig}
\end{figure}

The magnetic field ${\bf H}_{\rm probe}$ acting on the local magnetic probe has five different sources: 
exchange and dipole fields from the surrounding Cu spins, exchange and dipole fields from the surrounding Pr dipole moments, and the external magnetic field.
Contributions except the external magnetic field are directed along the $x$-direction, and we express their effect by the field $h_{x}$. We keep the field $h_{x}$ fixed as the direction of the external magnetic field ${\bf H}$ is changed which is the situation for small values of the magnetic field.\cite{janossy-1994}
For a general direction of the external magnetic field ${\bf H}$, we write the Hamiltonian of the local probe as
\begin{eqnarray}
{\cal H}_{\rm probe} = \hbar \gamma {\bf S}  \cdot {\bf H}_{\rm probe},
\end{eqnarray}
where $\gamma$ is the gyromagnetic ratio of the local probe, and ${\bf S}$ is its spin (either electron or nuclear).
Thus, we obtain the experimentally detected magnetic resonance shift, $h_{\rm probe}$ (in magnetic field units):
\begin{eqnarray}
 h_{\rm probe} = 
S (h_{\rm res} - h_{0}),
\end{eqnarray}
where $h_{\rm res}$ is the resonance field and $h_{0}$ is the resonance position for the AFI phase.
The external magnetic field is expressed by field angles as ${\bf H}/|{\bf H}|=[H_{x}, H_{y},H_{z}]=[\cos \phi \sin \theta, \sin \phi \sin \theta, \cos \theta]$, and we assume that $[S_{x}, S_{y}, S_{z}] \sim [H_{x}, H_{y},H_{z}]$. This gives
\begin{eqnarray}
h_{\rm probe} &=& (h_{x}+H\cos \phi \sin \theta)\cos \phi \sin \theta+H(\sin \phi \sin \theta)^2\nonumber\\
&+& H(\cos \theta)^2.
\end{eqnarray}
Figure~\ref{pheno-angle-fig} shows the field angle dependence of $h_{\rm probe}$ by changing the direction of the magnetic field in the planes $[001]$ and $[110]$.
We estimate the change of $h_{\rm probe}$ as
\begin{eqnarray}
\frac{h_{\rm probe}[100]-h_{\rm probe}[110]}{h_{\rm probe}[110]-h_{\rm probe}[001]} = \sqrt{2}-1,
\end{eqnarray}
which ratio can be checked in the NMR or ESR experiments.

\section{Discussion and Summary}

We studied the nature of the Pr ordered phase in PrBa$_{2}$Cu$_{3}$O$_{6}$ in a quasi-triplet CEF model of 4$f^{2}$ electrons, where we also included the symmetry allowed coupling between the Pr and Cu ions.
The reason to include Cu-Pr coupling is the Cu spin rotation observed at the Pr ordering temperature in neutron diffraction,\cite{boothroyd, lister-2001} which has not been observed for other rare earth ReBa$_{2}$Cu$_{3}$O$_{6}$ compounds.
In the 4$f$ rare-earth series, the localized electrons of Pr ion tend to hybridize rather strongly with conduction electrons as in e.g. PrFe$_{4}$P$_{12}$. This can be the reason why the Cu-Re coupling is much stronger in ReBa$_{2}$Cu$_{3}$O$_{6}$ for Re=Pr compared to the other rare-earth ions, Re.

We showed by a general symmetry analysis that there is no coupling between the Cu and Pr subsystems in phase AFI, but a coupling emerges in phase AFIII.
We prove that only the AFM ordering of $J_{x}(J_{y})$ dipole moments of Pr ions is consistent with the Cu spin rotation observed at $T=T_{\rm Pr}$  by neutron diffraction.\cite{boothroyd, lister-2001}
The interpretation of the neutron diffraction data in Ref.~\onlinecite{boothroyd} includes a tilting of the ordered Pr magnetic moments out of the $ab$-plane.
However, the presence of non-zero $J_{z}$ dipole component does not follow from the general symmetry analysis as it was shown in Section~\ref{cu-pr-interaction}.

Upon identifying of the Pr order parameter, we studied the Pr ordering temperature in the Cu-Pr coupled quasi-triplet CEF model by changing the interaction and CEF parameters. We found that the Cu-Pr coupling enhances the Pr ordering temperature, $T_{\rm Pr}$, compared to the uncoupled ordering temperature, $T_{0}$.

By fixing the Pr-Pr and Pr-Cu interaction parameters to the values obtained by neutron diffraction,\cite{lister-2001} and the CEF parameter to that obtained by high-temperature susceptibility measurements,\cite{uma} we found that the Pr ordering temperature is considerably enhanced due to the Cu-Pr coupling in the parameter range $\Delta \sim 8-10$\,K. Namely, the enhancement of the ordering temperature is as large as $T_{\rm Pr}/T_{0} \sim 8-12$ in this interval.
This observation explains the uniquely large ordering temperature for Re=Pr in the series ReBa$_{2}$Cu$_{3}$O$_{6+x}$.

In addition, we predicted the magnetic field angle dependence of the local magnetic field acting at a Pr site in the AFM ordered phase of the Pr ions.
This can be directly compared to measured NMR or ESR spectra which detect the local magnetic field by a local magnetic probe.

Our model contains the minimal number of parameters which is required to account quantitatively for the experimental data. Given that the parameter values are assumed from independent measurements and no fit is performed, the agreement is reasonable.
First, the realistic estimate $\Delta \sim 8-10$\,K gives a $T_{\rm Pr}$ of $5 - 3.6$\,K which is to be compared to the experimental $T_{\rm Pr}$ of $12$\,K. We note that this estimated range gives the Pr transition temperature $T_{0}$ for the uncoupled system as $0.7 - 0.3$\,K, which is close to the typical values of the ordering temperatures for other rare-earth ions such as Yb ($0.35$\,K), Nd ($0.5$\,K) or Dy ($1$\,K). This fact may also indicate that the uniquely large ordering temperature for Pr arises from the Cu-Pr coupling.
Second, the interval $\Delta \sim 8-10$\,K is not very far from the estimate $\Delta \approx 17$\,K based on the analysis of the high-temperature susceptibility data.\cite{uma}
Furthermore, the estimated interval for $\Delta$ and also the splitting pattern of the quasi-triplet subspace is consistent with the results of neutron scattering,\cite{hilscher-1994} which
found that the splitting of the ground state quasi-triplet does not exceed $2$\,meV above $T_{\rm Pr}$, and two inelastic lines are observed at energies $1.7$\,meV and $3.4$\,meV at $T=5$\,K in the ordered phase.
Finally, for the interval $\Delta \sim 8-10$\,K our CEF model gives the ordered Pr magnetic moment at low temperatures as $\mu_{\rm Pr}\sim 1.8 - 1.7\mu_{\rm B}$, which is to be compared to the observed value $\mu_{\rm Pr} = 1.15\mu_{\rm B}$.\cite{boothroyd}
The quantitative discrepancies between the results of our model and the experimental data arise because of the simplicity of our model and minimal number of parameters.

In summary, the model we propose herein describes the main features of PrBa$_{2}$Cu$_{3}$O$_{6}$ system, and
demonstrates the importance of the Cu-Pr coupling which may be the key point to understand the anomalous behavior of this compound such as the enhancement of the Pr ordering temperature. Further studies which include also the itinerant character of $4f^{2}$ electrons
are necessary to give an extensive description of this rather complicated system.

\acknowledgements

This work was supported by the Hungarian State OTKA Grant  K73455 and by the European Research Council Grant Nr. ERC-259374-Sylo.
AK acknowledges the Postdoctoral Programme of the Magyary Foundation and EGT Norway Grants.

%\bibliographystyle{apsrev}

%\bibliography{prbacuo}

\end{document}